\documentclass[10pt,preprintnumbers,nofootinbib,aps]{revtex4}
\usepackage{epsfig}

\def\Dirac#1{#1\hskip-5.5pt/ \hskip 1.pt}
\def\dslash{\Dirac\partial}
\def\dbar#1{\frac{d^4 #1}{(2\pi)^4}}

\begin{document}

\title{
Axial Vector Current and Coupling of the Quark in the Instanton Model }

\author{G. W. Carter}

\affiliation{
Department of Physics, University of Washington, Seattle, WA  98195-1560}

\date{27 August 2002}

\begin{abstract}
We compute the axial vector current, form factor, and coupling for quarks
in the instanton liquid model with two light flavors.
Non-local current corrections are derived, as required by the effective 
't Hooft interaction.
We obtain a pion-mediated axial form factor and an axial coupling which, 
when simply applied in the non-relativistic limit for constituent quarks, 
matches the experimental value to within a few percent, both in and out of 
the chiral limit.
\end{abstract}
\preprint{NT@UW-02-027}
\maketitle

\section{Introduction}

The axial coupling constant of the nucleon, deduced from measurements
of neutron beta decay, is known to be 
$| G_A/G_V | = 1.2670 \pm 0.0035$ \cite{PDG} in vacuum.
Strict chiral invariance would have this ratio be unity, and the substantial
difference suggests a deviation in excess of current quark mass effects.
Thus it has long been argued that the increase in $G_A$ is a 
result of the spontaneous chiral symmetry breaking that characterizes
the QCD vacuum.

Although additional chiral-invariant meson-nucleon couplings can remedy
this problem in the linear sigma model \cite{Lee}, a more fundamental
explanation is warranted.
The issue is complicated by the difficulties experienced by lattice
QCD practitioners \cite{BOS}, in that such calculations tend to yield
values significantly lower than experiment.
Possible explanations have recently centered on finite-size effects
\cite{Jaffe}, although the specific physics is unresolved \cite{Cohen}.

One might also try to deduce the nucleon's axial coupling by studying
that of a constituent quark, denoted in this paper as $g_A$.
In the non-relativistic limit,
\begin{equation}
G_A = \frac{5}{3} g_A \,,
\label{nrratio}
\end{equation}
but incomplete knowledge of the quark wave function leaves the
relativistic corrections unknown.
However there has been considerable success in treating the nucleon
as massive constituent quarks \cite{GellMann}, 
and if the deviation of $G_A$ from unity is in fact due to chiral forces the
constituent approach should encompass the relevant physics.
At the quark level chiral symmetry also demands an axial coupling of unity,
which generates an excess of nucleon axial charge when Eq.~(\ref{nrratio}) is
directly applied.
Studies addressing this predict negative corrections
of ${\cal O}(1)$ to $g_A$, using the MIT bag model \cite{CJJT},
in the large-$N_c$ limit \cite{Weinberg,
BSL}, or via perturbative corrections \cite{Duck}.

In this paper we consider corrections to $g_A$ from instantons.
The Instanton Liquid Model (ILM), in which the QCD vacuum is populated by
classical gauge configurations \cite{BPST,CDG}, 
dynamically generates the constituent quark mass
and has consistently led to phenomenological success when applied to problems 
related to spontaneous chiral symmetry breaking \cite{SS,Diakonov}.
As detailed below, we find a 25\% reduction in the axial coupling from
instantons, with only slight sensitivity to the chiral limit.
This is in excellent agreement with experiment if
corrections to Eq.~(\ref{nrratio}) are small.

To determine the axial coupling we will first carefully analyze the axial
current.
Since instanton-induced interactions between quarks are non-local, the
currents associated with conserved quantities are non-trivial.
Next, given the necessity of pions in axial phenomena, meson correlation 
functions are briefly reviewed.
We then derive the axial current and verify the importance of the pion 
pole in the axial form factor, $h_A(q^2)$.
Finally, the leading corrections to the coupling $g_A$ are computed, taking
into account multi-instanton effects and substantial
non-local vertex corrections.
In the chiral limit we find $g_A = 0.73$ and estimate
that with a physical pion mass $g_A = 0.77$.
With naive application of Eq.~(\ref{nrratio}), these correspond to a
nucleon axial couping of 1.22 and 1.28, respectively.

\section{Non-Locality and Conserved Currents}

The effects of an instanton liquid background are most economically encoded
in an effective quark action \cite{DP2}.
Taking the exact zero-mode solutions to the Dirac equation 
and isolating them as the dominant low-energy effects leads to a 
$2 N_f$-quark vertex function \cite{tHooft}, 
similar to the Nambu--Jona-Lasinio model of QCD \cite{NJL,VL,Klevansky}.
A well-studied simplification of this model, chiral Random Matrix Theory, has 
shown that chiral symmetry breaking via random overlap integrals 
reproduces Dirac eigenvalue correlations as computed on the lattice
\cite{BMSVW}.
Thus such an approach seems to indeed contain the essence of
non-perturbative QCD.
Furthermore, the ILM relies on only two parameters: the instanton density
and average instanton size.
Both of these were phenomenologically fixed long ago with the vacuum
energy density and chiral condensate 
\cite{Shuryak,DP2} and measured on the lattice \cite{HN}.
These parameters determine the diluteness of the instanton liquid, the
ratio of the average size to inter-instanton spacing, to be about 1/3, a
somewhat small parameter which allows for perturbative treatment of the
instanton-induced vertex.

The Eulcidean effective action for $N_f$ flavors in the chiral 
limit is written \cite{Diakonov},
\begin{equation}
S = -\int\!d^4x \; \psi^{\dagger}(x) i\dslash \psi(x)
+ i \lambda \int \! dU\;d^4z\;\prod_f^{N_f} \Big[ - d^4x_f\;d^4y_f\;
\psi_f^{\dagger}(x_f) i\dslash \Phi(x_f-z,U) \tilde\Phi(y_f-z,U) i\dslash
\psi^f(y_f)\Big], 
\label{oldsint} 
\end{equation}
where $U$ is the
instanton's $2\times N_c$ color/spin orientation matrix and $z$ is its
position. 
The $\Phi(x)$ is the zero mode solution for fermions in the
field of one instanton, and its Fourier transform is the form factor,
\begin{equation} 
f(p) = 2 x \left[
I_1(x)K_0(x)-I_0(x)K_1(x)+\frac{1}{x}I_1(x)K_1(x) \right]_{x=p\rho/2}\,,
\label{ffac} 
\end{equation} 
which provides a natural momentum cutoff of scale $\rho^{-1}$.
We have left out additional (quarkless) $\lambda$ terms which relate 
this coupling
constant, which is in fact a Lagrange multiplier, to the instanton density,
$N/V$.

Because of non-locality Eq.\,(\ref{oldsint}) is not invariant under 
symmetry transformations.
Since bulk observables are often insensitive to this problem it is 
usually ignored, however when addressing currents the dependence is crucial.
A literature exists in which non-local interactions are modified to be 
invariant \cite{Schwinger,Terning,BKN,BB,PB, GBR,ADT} 
and we follow the same procedure here. 
Taking the non-local four-fermion interaction as the starting point, it
then becomes a matter of multiplying each quark operator by a
path-ordered exponential in the background of a source gauge field
$a_\mu$, replacing the Euclidean quark fields as:
\begin{eqnarray} 
\psi(x) &\rightarrow& W(z,x)\psi(x), \nonumber\\ 
\psi^{\dagger}(x) &\rightarrow& \psi^{\dagger}(x)\gamma_0 W(x,z)
\gamma_0  \,, \nonumber
\end{eqnarray} 
with the operator \cite{Schwinger}
\[
W(x,y) = {\cal P}{\rm exp} \left(
- i\int^y_x\! ds_{\mu} a^a_{\mu} T^a\right).
\]
As has been pointed out in the cited works, the operators $W(x,y)$ must be
path-ordered and the choice of path integrated over is not unique. 
However, we are primarily concerned with the longitudinal currents,
for which results are independent of any particular choice of path \cite{PB}.

Thus we write the modified action as
\begin{eqnarray}
S &=& -\int\!d^4x \; \psi^{\dagger}(x) i\dslash \psi(x)
\nonumber\\ &&
+ i \lambda \int dU\;d^4z\;\prod_f^{N_f} \Big[- d^4x_f\;d^4y_f\;
\psi_f^{\dagger}(x_f) 
\gamma_0 W(x_f,z) \gamma_0 i\dslash \Phi(x_f-z) \Phi^{\dagger}(y_f-z)
i\dslash W(z,y_f) \psi^f(y_f)\Big] .
\label{sint} 
\end{eqnarray}
We now concentrate on the case of two light quarks, $N_f=2$.

The Noether currents can be evaluated directly from the action.
In momentum space,
\[
j_\mu^a(q) = \left.-\frac{\delta S}{\delta a_\mu^a(q)}\right|_{a_\mu^a=0} .
\]
We thus have, after Fourier transforming the fields and averaging over colors,
\begin{equation}
j_\mu(q) = \int \dbar{p}\, \psi^{\dagger}(p) \gamma_\mu T \psi(p+q) 
+ j_\mu^L(q) + j_\mu^R(q),
\end{equation}
where the left-handed non-local term is
\begin{eqnarray}
j_\mu^L(q) &=& 
 \frac{\lambda(2\pi\rho)^4}{N_c^2-1} \epsilon^{f_1f_2} \epsilon_{g_1g_2} 
\left(\delta^{i_1}_{j_1}\delta^{i_2}_{j_2} - \frac{1}{N_c}
\delta^{i_1}_{j_2}\delta^{i_2}_{j_1}\right)
\int
d^4z d^4x \dbar{p_1}\dbar{p_2}\dbar{k_1}\dbar{k_2}\dbar{p'} 
\nonumber\\ &&
\times \int_x^z ds_\mu e^{-iq\cdot s}
\Big\{ e^{-ix\cdot(p_1-p')-iz\cdot(p'+p_2-k_1-k_2) }
  \left[\psi_L^\dagger(p_1) 
  \gamma_0T\gamma_0\right]_{f_1i_1}\psi_L^{g_1j_1}(k_1) f(p')f(k_1)
\nonumber\\ && \quad
- e^{ix\cdot(k_1-p')-iz\cdot(p_1+p_2-p'-k_2) }
  \psi^\dagger_{Lf_1i_1}(p_1)\left[T \psi_L(k_1)\right]^{g_1j_1}
  f(p_1)f(p')
\Big\}
\psi^\dagger_{Lf_2i_2}(p_2) \psi_L^{g_2j_2}(k_2) f(p_2) f(k_2) \,,
\end{eqnarray}
and the right-handed $j_\mu^R(q)$ is similar defined.
Group indices have been suppressed; note that these are carried
by the operators $T$.
The instanton or anti-instanton at the core of each vertex naturally
splits the quark fields into right- or left-handed spinors,
$\psi_{R/L} = \frac{1}{2}(1\pm\gamma_5)\psi$.
Since we are only concerned here with spin structures of the identity 
matrix or $\gamma_5$, the decomposition is trivial.
Carrying out the spatial integrations, we have
\begin{eqnarray}
&&j_\mu^L(q) =
\frac{q_\mu}{q^2}
\frac{i\lambda(2\pi\rho)^4}{N_c^2-1} \epsilon^{f_1f_2} \epsilon_{g_1g_2} 
\left(\delta^{i_1}_{j_1}\delta^{i_2}_{j_2} - \frac{1}{N_c}
\delta^{i_1}_{j_2}\delta^{i_2}_{j_1}\right)
\int \dbar{p_1}\dbar{p_2}\dbar{k_1}\dbar{k_2}
\delta(p_1+p_2-k_1-k_2+q)
\nonumber\\ && \quad \times
\left\{ \left[\psi_L^\dagger(p_1) \gamma_0T\gamma_0\right]_{f_1i_1}
\psi_L^{g_1j_1} (k_1) \left[f(p_1)-f(p_1+q)\right] f(k_1) - 
\psi^{\dagger}_{Lf_1i_1}(p_1)
\left[T \psi_L(k_1)\right]^{g_1j_1} f(p_1)\left[f(k_1)-f(k_1-q)\right] \right\}
\nonumber\\ && \quad \times
\psi^{\dagger}_{Lf_2i_2}(p_2)\psi_L^{g_2j_2}(k_2) f(p_2)f(k_2) 
\, .
\label{currl}
\end{eqnarray}

In the event of spontaneous symmetry breaking, quark operators which 
overlap with the condensate can be paired into loops \cite{BB,CD2}.
When chiral symmetry is broken an effective quark mass is generated,
dependent on four-momentum $p=\sqrt{p_\mu p_\mu}$ as
\[
M(p) = M f(p)^2\,.
\]
The constant $M$ is the solution to a gap equation 
obtained by minimization with respect to the coupling $\lambda$ \cite{DP1},
\begin{equation}
\frac{N}{V} = 4 N_c \int \dbar{p} \frac{M(p)^2}{p^2+M(p)^2}\,,
\label{gapeq}
\end{equation}
and the constant $N/V = (200\,{\rm MeV})^4$ is the instanton density.
With the phenomenological $\rho = 1/3$ fm, one finds $M = 345$ MeV.
In this case the current, Eq.~(\ref{currl}), reduces to
\begin{eqnarray}
&&j_\mu^L(q) =
-i\frac{q_\mu}{q^2} \int \dbar{p}\, \psi^\dagger_L(p) 
\Bigg\{ M(p+q) \gamma_0T\gamma_0 - M(p) T
\nonumber\\ &&  \quad
  + \frac{\lambda(2\pi\rho)^4}{N_c} f(p) f(p+q) \int \dbar{k}
    f(k) \frac{M(k)}{k^2+M(k)^2} {\rm Tr} \left[ f(k+q) 
    \frac{1-\gamma_5}{2}\gamma_0T
    \gamma_0 - f(k-q) T 
    \frac{1-\gamma_5}{2} \right] \Bigg\} \psi_L(p+q) 
,
\label{gencurr}
\end{eqnarray}
where the trace is over spin only.

Further reduction of this still-unwieldy expression requires specifying
a particular current and associated group elements $T$.
The simplest example is the quark number current, $B_\mu$, where $T=\openone$:
\begin{equation}
B_\mu(q) = \int\dbar{p}\, \psi^\dagger(p) \left\{\gamma_\mu 
 - i\frac{q_\mu}{q^2} \left[ M(p+q) - M(p)\right] \right\}
\psi(p+q) \,.
\label{qnum}
\end{equation}
As $q\rightarrow 0$, this becomes \cite{CD1,GBR}
\[
B_\mu(0) = \int\dbar{p} \,
 \psi^\dagger(p) \left( \gamma_\mu  - i \frac{d M(p)}{dp_\mu} \right)
\psi(p) \,.
\]
It is easy now to see that the $q$-dependent term in Eq.~(\ref{qnum}), 
a result of the non-locality of the interaction, 
is necessary for current conservation.
With Eq.~(\ref{qnum}) and the Dirac equation,
\begin{equation}
\left[ \Dirac{p} - iM(p)\right]\psi(p) = 0 \,,
\label{diraceqn}
\end{equation}
we see the standard cancellation ensues and 
\begin{equation}
q_\mu B_\mu(q) = 0 \,.
\end{equation}
This is the simplest illustration of the more general rule, which becomes
more complicated when the axial isovector current is considered.
But before analyzing this, we review the pion's derivation in 
the ILM.

\section{\label{mesons}Meson Correlation Functions}

\begin{figure}[tb]
\epsfig{file=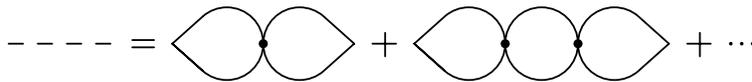, width=104mm}
\caption{
The series of instanton ladder diagrams summed to obtain meson correlation
functions.
}\label{pi_fig}
\end{figure}

The pion's role in axial charge currents has been long established.
To be made explicit in the ILM, one must sum all ladder diagrams using
the 't Hooft vertex, a series shown in Fig.~\ref{pi_fig}.
This was originally done in Ref.~\cite{DP3}, although the advent of the
effective action (\ref{sint}) has since made this procedure simpler.

In the chiral limit, one finds the expected pole in the pion propagator
as $q^2\rightarrow 0$.
Furthermore, the use of non-local corrections to the axial currents has 
been shown to produce the required transversality in the mesonic
current and properly relate the pion decay and renormalization
constants \cite{BB}.
Following the results of these cited works,
we have the pion correlation function of Fig.~\ref{pi_fig},
\begin{eqnarray}
\Pi_\pi^{AB}(q^2) &=& \int\dbar{p} \langle \left( \psi^\dagger(p) 
\tau^A \gamma_5 \psi(p+q) \right)
\left( \psi^\dagger(p+q) \tau^B \gamma_5 \psi(p) \right) \rangle 
\nonumber\\ &=&
-\frac{\delta^{AB}}{4 N_c^2} \left( \frac{1}{i\lambda} -
\frac{2}{N_c M^2}\int \dbar{p} 
\frac{ M(p)M(p+q) \left[ p\cdot(p+q) + M(p)M(p+q)\right]}
{\left(p^2+M(p)^2\right)\left[(p+q)^2+M(p+q)^2\right] } \right)^{-1} .
\end{eqnarray}
The gap equation (\ref{gapeq}) can be written \cite{CD1}
\begin{equation}
\frac{1}{i\lambda} = \frac{2}{N_cM^2}\int\dbar{p}\frac{M(p)^2}{p^2+M(p)^2},
\end{equation}
and one finds, upon expanding in small $q^2$,
\begin{eqnarray}
\Pi_\pi^{AB}(q^2) &=& - \frac{\delta^{AB}}{4 N_c q^2}
\left( \frac{1}{M^2}\int\dbar{p}\frac{M(p)^2-\frac{1}{2}pM(p)M'(p)
+\frac{1}{4}p^2M'(p)^2 }{\left[p^2+M(p)^2\right]^2 }\right)^{-1}
\nonumber\\
&\equiv& -\frac{g_{\pi qq}^2}{q^2}\delta^{AB} .
\end{eqnarray}
The coupling constant is determined thusly by the compositeness 
condition \cite{Weinberg2} and,
with the standard instanton parameters listed above, we find
$g_{\pi qq} = 3.78$.
The pion and other channels were analyzed in an NJL model with alternative
form factors 
by Plant and Birse, who found $g_{\pi qq} = 3.44$ for the pion and the 
sigma parameters $g_{\sigma q q}= 3.51$ and  
$m_\sigma = 443$ MeV \cite{PB}. 
This required extracting the sigma pole, and instead of repeating this
procedure or fitting the sigma correlation function \cite{Hutter} we will
simply scale our pion-quark coupling in a similar manner to obtain
$g_{\sigma q q}= 3.86$.
These two channels will be sufficient in the following analysis of the
axial properties of the quark.

\section{Axial Current and Form Factor}

We now address the axial isovector current.  
A key experimental quantity is the matrix element of the current,
$A_\mu^a(p_2-p_1)$, taken with initial and final state nucleons of 
momenta $p_1$ and $p_2$.
Here we compute the analogous quantity with quark fields:
\begin{equation}
\langle \psi(p_2) \vert A_\mu^a(q) \vert \psi(p_1) \rangle =
u^\dagger(p_2) \frac{\tau^a}{2}\left[ \gamma_\mu\gamma_5 g_A(q^2) 
+ q_\mu \gamma_5 h_A(q^2)
\right] u(p_1) \,, 
\label{axialdecomp}
\end{equation}
where the $u^\dagger(p_2)$ and $u(p_1)$ are the Euclidean spinor solutions 
of the Dirac equation for free quarks, and $q=p_2-p_1$.
We are ultimately interested in the limit of $p_1=p_2=0$, or on-shell
quarks at zero momentum transfer, with
the constant $g_A = g_A(0)$ the axial coupling. 
First we consider $h_A(q^2)$, the axial form factor.
With a nonzero $g_A$ and current conservation, it is clear in the chiral 
limit that $h_A(q^2)$ has a pole at $q^2=0$, identified as the pion.
Through explicit construction, we will show this pole is manifest in the 
ILM.

\begin{figure}[tb]
\epsfig{file=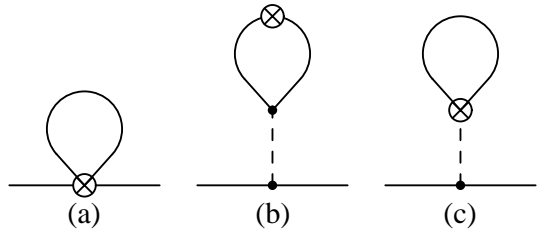, width=80mm}
\caption{
Diagrams contributing to the axial form factor, $h_A(q^2)$.  Circled 
crosses are current insertions and the dashed line denotes a pion.
}\label{ha_fig}
\end{figure}

Inserting
$T = \gamma_5\tau/2$ into Eq.~(\ref{gencurr}), 
we have bare and single instanton (and anti-instanton) contributions in
\begin{eqnarray}
A^a_{\mu}(q)_{\ref{ha_fig}a} 
&=& \int\dbar{p}\,\psi^\dagger(p) \frac{\tau^a}{2}\gamma_\mu\gamma_5
\psi(p+q) + i \frac{q_\mu}{q^2}\int\dbar{p}\,\psi^\dagger(p)\frac{\tau^a}{2}
\gamma_5 \Bigg\{ M(p+q)+M(p) 
\nonumber\\ &&
- \frac{2i\lambda(2\pi\rho)^4}{N_c}
f(p)f(p+q)\int\dbar{k} f(k) \left[f(k+q)+f(k-q)\right]
\frac{M(k)}{k^2+M(k^2)} \Bigg\} \psi(p+q) \,.
\label{axone}
\end{eqnarray}
Graphically, this is Fig.~\ref{ha_fig}a.

Clearly, no amount of labor will reveal this to be a conserved current;
as the appearance of the chiral Goldstone boson is a multi-instanton effect, 
so too is axial vector current conservation.
Pion tadpoles, as rendered in Figs.~\ref{ha_fig}b and \ref{ha_fig}c,
are necessary to remove the finite divergence of $A_\mu^a$.
The first is a bare current insertion, the second the non-local piece.
Their sum is
\begin{equation}
A^a_{\mu}(q)_{\ref{ha_fig}b+\ref{ha_fig}c}
= -\frac{q_\mu}{q^2}\frac{2\lambda(2\pi\rho)^4}{N_c} 
\int\dbar{p}  f(p)f(p+q) \,
\psi^\dagger(p) \frac{\tau^a}{2}\gamma_5 \psi(p+q)
\int\dbar{k} f(k)\left[f(k+q)+f(k-q)\right]\frac{M(k)}{k^2+M(k)^2} \,.
\end{equation}
This cancels the final term in Eq.~(\ref{axone}) and the full current is
\begin{equation}
A_\mu^a(q) = \int\dbar{p}\,\psi^\dagger(p)
\left\{ \gamma_\mu
+ \frac{q_\mu}{q^2} \left[ iM(p)+iM(p+q) \right]\right\} 
\gamma_5\frac{\tau^a}{2} \psi(p+q) \,.
\label{axcurrent}
\end{equation}
The first term is the bare axial coupling, unity with the definition
of Eq.~(\ref{axialdecomp}), and the second is 
the pion pole carried by the axial form factor,
\begin{equation}
h_A(q^2) = -i\frac{M(p+q)+M(p)}{q^2} \,.
\end{equation}
Note that at long wavelength the leading divergent part of the form factor 
is the non-local,
quark level version of the classic nucleon result, $h_A(q^2)= -2iM_N/q^2$.
The pion pole similarly couples separately to each constituent quark.

It is now clear, with the Dirac equation (\ref{diraceqn}), that
\begin{equation}
q_\mu A_\mu^a(q) = 0 \,.
\end{equation}
Equivalently, with the propagator,
\[
S(p)= \frac{\Dirac{p}+iM(p)}{p^2+M(p)^2} \,,
\]
the Ward-Takahashi Identity is satisfied,
\begin{equation}
q_\mu \Gamma_{\mu 5}^a = - \frac{\tau^a}{2} \left[
\gamma_5 S(p+q)^{-1} + S(p)^{-1}\gamma_5 \right].
\end{equation}

\section{Axial Coupling Constant}

The axial form factor involved the effects of multiple instantons, with
pions transferring $t$-channel momentum.
Similar propagation in the $s$-channel contributes to the
axial coupling constant.
While still dominant, the pion is not the only relevant resonance in 
this channel.
Since the sigma's coupling constant is of comparable size, and its mass is
only slightly greater than that of a constituent quark, its effects will
also be taken into account.
Higher resonances, such as the $\rho$ and $A_1$ vectors, are both substantially
heavier and have quark-meson couplings a mere third of the scalar and 
pseudoscalar \cite{PB,Hutter}.
With the leading contributions of order $g_{iqq}^2$, these are safely
ignored.

\begin{figure}[tb]
\epsfig{file=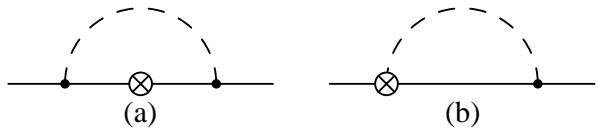, width=83mm}
\caption{
Diagrams contributing to the axial coupling constant, $g_A$.  Circled 
crosses are current insertions and dashed lines denote a either
a pion or sigma meson propagator.
}\label{ga_fig}
\end{figure}

In Eq.~(\ref{axcurrent}) we have seen that the bare
$g_A$ is unity.
To this we first add the pion contribution of Fig.~\ref{ga_fig}a, which is
\begin{equation}
g_A^{\ref{ga_fig}a,\pi} = 
- \frac{g_{\pi q q}^2 }{2M} \int\dbar{p} \frac{M(p)}{p^2}
\frac{p^2+2 M(p)^2}{[p^2+M(p)^2]^2} \,.
\label{pion1}
\end{equation}
After evaluation, the identical sigma contribution differs only
by the coupling constant and massive propagator in the integrand,
\begin{equation}
g_A^{\ref{ga_fig}a,\sigma} = 
- \frac{g_{\sigma q q}^2 }{2M} \int\dbar{p} \frac{M(p)}{p^2+m_\sigma^2}
\frac{p^2+2 M(p)^2}{[p^2+M(p)^2]^2} \,.
\label{sigma1}
\end{equation}
These expressions evaluate to $-0.102$ and $-0.0425$, 
respectively when we take the parameters described in Section \ref{mesons}.

The non-local contributions couple the axial current directly to the 
four-quark vertex, as shown in Fig.~\ref{ga_fig}c.
With an internal pion, we find
\begin{equation}
g_A^{\ref{ga_fig}b,\pi} = 
\frac{3 g_{\pi q q}^2 }{4M} \int\dbar{p} \frac{M'(p)} {p[p^2+M(p)^2]}\,,
\end{equation}
and for the sigma,
\begin{equation}
g_A^{\ref{ga_fig}b,\sigma} = 
\frac{g_{\sigma q q}^2 }{4M} \int\dbar{p} \frac{pM'(p)} 
{(p^2+m_\sigma^2)[p^2+M(p)^2]}\,.
\end{equation}
Numerically, we obtain $-0.101$ and $-0.0236$ for these corrections.
Note that both numbers are comparable to those of Fig.~\ref{ga_fig}a, 
{\em i.e.} the non-local corrections are not suppressed in any systematic
way.

In the chiral limit, we therefore find
\[
g_A = 1 - 0.102 -0.0425 - 0.101 - 0.0236 = 0.73 \,,
\]
in excellent agreement with the non-relativistically deduced value.

The vector current, to which the axial coupling is compared experimentally,
receives no contributions from the non-local vertices.
Furthermore, a critical sign difference in the numerator as compared 
to Eqs.~(\ref{pion1}) and (\ref{sigma1}) leads to a finite but very small 
contribution
from each diagram, summing to only two percent of the bare coupling.
Thus we can ignore such instanton effects in the coupling $g_V$ and
take it to be unity.

In order to estimate the correction from finite current quark masses,
we assume larger masses for the internal meson propagators.
With the physical pion mass, $m_\pi = 138$ MeV, the instanton
calculations of Hutter \cite{Hutter} generate a sigma mass of
$m_\sigma = 540$ MeV.
Reevaluating, we naturally find a deduction in these subtractions,
\[
g_A = 1 - 0.079 -0.036 - 0.094 - 0.021 = 0.77 \,.
\]
However, this is a mere 5\% increase compared to the chiral limit, and 
in fact places the result extremely close to the pseudo-experimental value.

We are ultimately interested in the nucleon axial coupling, and in 
the non-relativistic limit we find
\begin{equation}
G_A = \frac{5}{3} g_A = \left\{
\begin{array}{cl}
1.22 & \mbox{for $m_\pi=0$} \\
1.28 & \mbox{for $m_\pi=140$ MeV} 
\end{array}\right. .
\end{equation}
The physical value, $G_A = 1.2670$ \cite{PDG}\footnote{
Our convention differs from that reported by the Particle Data Group
by an overall sign.
}, lies in between.

\section{Conclusions}

Through an analysis of current conservation in the instanton liquid model,
we found that terms arising from the non-locality of the 't Hooft vertex
were central to maintaining the symmetries of the theory.
This can be thought of as a correction for the truncation of quark basis
states, for we have isolated the zero modes and removed them from 
the dynamics.
Reintroducing the symmetries to the volume of this extended interaction
vertex via path-ordered exponentials recovers the symmetries, as has been
noted by other authors some time ago \cite{Terning,BKN}.
A general form of current corrections has been computed here, followed
by specialization to the axial iso-vector case.

After these modifications the pion pole appears as a $t$-channel 
quark-antiquark correlation, governing each constituent quark's axial 
form factor and maintaining current conservation.
Similar multi-instanton corrections, when applied with $s$-channel
chiral mesons, lead to an axial coupling constant of the quark of 
$g_A = 0.73$.
In the non-relativistic limit this
corresponds to a nucleon coupling of $G_A = 1.22$, within four percent
of the experimental value.
Away from the chiral limit, when the pion is set to its physical mass,
agreement improves to within one percent.
These numbers are obtained with the two standard 
inputs of the instanton vacuum, the number density and average size, 
and no additional parameters.
While there are additional corrections from vector mesons, the significantly
lower coupling constants and higher masses render them negligible.
However, the non-local current terms lead to contributions comparable to 
the local ones and cannot be disregarded.

Our results suggest that the ILM contains the primary
effects which lead to the observed vacuum axial coupling of 1.267.
Given that this variance from unity is a likely result of spontaneous
chiral symmetry breaking, a central result of the ILM, it is not surprising 
that instantons contribute favorably.
It is, however, noteworthy in its implication that relativistic
wave function corrections to the constituent quark are nearly negligible.

\begin{acknowledgments}
I thank T. Sch\"afer and G. Miller for useful conversations.
This work was supported by US-DOE grant DE-FG03-97ER4014.
\end{acknowledgments}

\end{document}